\documentclass[twocolumn,showpacs,preprintnumbers,amsmath,amssymb,superscriptaddress]{revtex4}

\usepackage{graphicx}
\usepackage{dcolumn}
\usepackage{bm}
\begin{document}

\title{Landau-Zener-St\"{u}ckelberg Interference of Microwave Dressed States
of a Superconducting Phase Qubit}

\author{Guozhu Sun}
\email{gzsun@nju.edu.cn}
\affiliation{Research Institute of Superconductor Electronics,School of Electronic
Science and Engineering, Nanjing University, Nanjing 210093, China}
\affiliation{Department of Physics and Astronomy, University of Kansas, Lawrence, KS
66045, USA}
\affiliation{National Laboratory of Solid State Microstructures, School of Physics, Nanjing University, Nanjing
210093, China}
\author{Xueda Wen}
\affiliation{National Laboratory of Solid State Microstructures, School of Physics,
Nanjing University, Nanjing 210093, China}
\author{Bo Mao}
\affiliation{Department of Physics and Astronomy, University of Kansas, Lawrence, KS
66045, USA}
\author{Yang Yu}
\affiliation{National Laboratory of Solid State Microstructures, School of Physics,
Nanjing University, Nanjing 210093, China}
\author{Jian Chen}
\affiliation{Research Institute of Superconductor Electronics,School of Electronic
Science and Engineering, Nanjing University, Nanjing 210093, China}
\affiliation{National Laboratory of Solid State Microstructures, School of Physics, Nanjing University, Nanjing
210093, China}
\author{Weiwei Xu}
\affiliation{Research Institute of Superconductor Electronics,School of Electronic
Science and Engineering, Nanjing University, Nanjing 210093, China}
\affiliation{National Laboratory of Solid State Microstructures, School of Physics, Nanjing University, Nanjing
210093, China}
\author{Lin Kang}
\affiliation{Research Institute of Superconductor Electronics,School of Electronic
Science and Engineering, Nanjing University, Nanjing 210093, China}
\affiliation{National Laboratory of Solid State Microstructures, School of Physics, Nanjing University, Nanjing
210093, China}
\author{Peiheng Wu}
\affiliation{Research Institute of Superconductor Electronics,School of Electronic
Science and Engineering, Nanjing University, Nanjing 210093, China}
\affiliation{National Laboratory of Solid State Microstructures, School of Physics, Nanjing University, Nanjing
210093, China}
\author{Siyuan Han}
\email{han@ku.edu}
\affiliation{Department of Physics and Astronomy, University of Kansas, Lawrence, KS
66045, USA}

\begin{abstract}
We present the first observation of Landau-Zener-St\"{u}ckelberg (LZS)
interference of the dressed states arising from an artificial atom, a superconducting phase
qubit, interacting with a microwave field.
The dependence of  LZS interference fringes on various external parameters and the initial state of the qubit agrees quantitatively very well with the theoretical prediction. Such LZS interferometry between the dressed states enables us to control the quantum states of a tetrapartite solid-state system with ease, demonstrating the feasibility of implementing efficient multipartite quantum logic gates with this unique approach.
\end{abstract}

\maketitle

The energy level diagram of quantum systems, such as atoms and nuclear spins, may
exhibit avoided level crossings as a function of an external control parameter, as shown in the inset
of Fig. 1(a). If one varies the
external control parameter to sweep the system across one of the avoided level crossings
back and forth, the quantum states evolving along the two
different paths will interfere, generating the well-known Landau-Zener-St%
\"{u}ckelberg (LZS) interference \cite{Shevchenko2010}, which was
originally observed in helium Rydberg atoms \cite{PhysRevLett.69.1919}.
Recent progress in solid-state qubits has stimulated strong interests in LZS
interference in superconducting qubits \cite%
{AnnalsofPhysics,WilliamD.Oliver12092005,PhysRevLett.96.187002,PhysRevLett.98.257003,
PhysRevLett.101.017003,PhysRevLett.101.190502,sun:102502,Nature.459.960,
Shevchenko2010,Nat.Commun.Sun} and other systems \cite{Garg,J.R.Petta02052010}. However,
most of  the previous work was performed in simple systems having avoided level crossings in their energy
diagrams. Since monochromatic electromagnetic fields have been extensively used to
control quantum states, for both theoretical curiosity and practical significance it is interesting and important
to know whether LZS interference can be realized and observed between the dressed states \cite{API}, generated from the interaction between photons
and atoms or even macroscopic quantum objects such as superconducting
qubits \cite{RevModPhys.73.357,PhysicsToday.58.11,Nature.453.1031}. The latter interaction is opening a new field named circuit quantum
electrodynamics (C-QED) \cite{PhysRevA.67.042311,PhysRevB.68.064509,Nature451.664}.
Although creating avoided crossings with dressed states of a Cooper pair box has been proposed \cite%
{PhysRevLett.98.257003,PhysRevB.81.024520}, no evidence of LZS
interference has been reported so far.

In this letter, we report the first
observation of LZS interference of the microwave dressed states of a
superconducting phase qubit (SPQ) \cite{QuantumInfProcess.8.81} by using nanosecond triangle pulses to sweep the system across
the avoided crossing between the microwave dressed qubit states. We show that the observed oscillations in
the SPQ's occupational probability are the result of
LZS interference. Furthermore, we develop a theoretical model based on the microwave
dressed states that quantitatively reproduces the dependence of LZS interference fringes on the sweep rate,
the microwave power, the microwave frequency,
and the initial state of the qubit. Since these external parameters can be controlled precisely in the experiments,
LZS interferometry of the microwave dressed states may provide a new approach to improving the speed and
fidelity of quantum information processing.

One form of the SPQ is based on the rf-SQUID, which
consists of a superconducting loop interrupted by a Josephson junction as
shown in Fig. 1(b). The superconducting phase difference $\varphi $ across the
junction serves as this macroscopic quantum object's dynamic variable. Such ``phase particle''
has a discrete eigenenergy spectrum which is a function of the external flux
bias. When properly biased, the ground and first excited states in one
of the potential wells act as $|0\rangle $ and $|1\rangle $ of the qubit \cite{QuantumInfProcess.8.81},
respectively. Fig. 1(a) shows the measured spectroscopy of the SPQ
used in the experiments. Setting $\hbar =1$, the level spacing between $%
|1\rangle $ and $|0\rangle ,$ $\omega _{10}=\omega _{1}-\omega _{0},$
decreases with the external flux bias due to the anharmonicity of the
potential well. Ideally, one expects that $\omega _{10}$ would be a continuous
function of the flux bias. However, three avoided crossings with splittings 2$g_1/2\pi$= 60 MHz, 2$g_2/2\pi$ = 22 MHz and 2$g_3/2\pi$ = 46 MHz, near 16.45 GHz,
16.21 GHz, and 16.10 GHz, respectively, resulting from the couplings between the qubit and
microscopic two-level systems (TLSs) \cite%
{QuantumInfProcess.8.81,QuantumInfProcess.8.117} were observed.
Although the microscopic origin of TLSs and the mechanism of their interaction with qubits
are still unclear and difficult to control, their quantum nature has been explored for quantum information
applications such as quantum memory \cite{NaturePhysics.4.523} and
qubit \cite{PhysRevLett.97.077001,Nat.Commun.Sun}.

The experimental procedure to realize and observe LZS interference between the microwave dressed qubit states
is depicted in Fig. 1(c). The qubit initialized in $|0\rangle $ and dc
biased at $\Phi _{i}$. Then a microwave pulse of width $t_{mw}$ was
applied, which generated a set of dressed states. The microwave frequency $%
\omega $ was chosen to be greater than $\omega _{10}$ at $\Phi _{i}$.
The intersecting of the dressed states $|0,n+1\rangle $ and $|1,n\rangle$
produced an avoided crossing. At the same time, a concurrent triangle pulse of width
$t_{\Lambda}$ = $t_{mw}$ was used to sweep the system's instantaneous flux bias from $\Phi _{i}$
to $\Phi _{LZS}$ and then back to $\Phi _{i}$. After turning off the
triangle pulse, the population of the qubit state $|1\rangle $, $P_{1}$, was
measured. Then we repeated the above process with different values of $\Phi
_{LZS}$ and $t_{\Lambda}$ to obtain a plot of $P_{1}$ versus $\Phi
_{LZS}$ and $t_{\Lambda}$ as shown in Fig. 2(a). (Note that in
Fig. 2 and Fig. 3, $\Phi _{LZS}$ is measured with respect to $\Phi _{i}$.) For $\Phi _{LZS}<\Phi _{r}$,
where the microwave was resonant with $\omega _{10}$, the
system could not reach the avoided crossing,  and thus there was no Landau-Zener (LZ)
transition and LZS interference. When the amplitude
of the triangle pulse was increased to $\Phi _{LZS}>\Phi _{r}$, striking interference
fringes appeared. The positions of these interference fringes in the $\Phi
_{LZS}-t_{\Lambda}$ plane were nearly independent of the microwave power (Fig. 2(b)
and Fig. 2(c)) but dependent on the microwave frequency (Fig. 2(d)). When $\omega $
decreased, $\Phi _{r}$ moved closer to $\Phi _{i}$ as expected according
to the measured spectrum shown in Fig. 1(a). The resulting LZS interference
fringes also moved closer to $\Phi _{i}$.

Using the dressed states picture, we can readily capture the underlying
physics and provide a quantitative description of the observed interference patterns. The
Hamiltonian of the microwave-dressed qubit can be written as $
H_{0}=H_{q}(t)+H_{m}+H_{q-m}$. $H_{q}(t)=\frac{1}{2}\omega _{10}(t)\sigma
_{z}^{q}$ is the Hamiltonian of the qubit, where $\sigma _{z}^{q}$ is Pauli Z operator on the qubit and $\omega _{10}(t)$ is the
energy level spacing of the bare qubit, which can be controlled \emph{in situ}
by the triangular pulse. The Hamiltonian of the microwave field is $%
H_{m}=\omega a^{\dagger }a$, where $a^{\dagger }$ and $a$ are the creation
and annihilation operators, respectively. The interaction Hamiltonian then
is $H_{q-m}=g(a^{\dagger }\sigma _{-}^{q}+a\sigma _{+}^{q})$, where $g$ is
the coupling strength between the microwave field and the qubit, $\sigma _{-}^{q}$ and $\sigma _{+}^{q}$
are the raising and lowering operators on the qubit. Truncating $%
H_{0}$ in the subspace spanned by $\{|1,n\rangle ,|0,n+1\rangle \}$, where $%
|n\rangle $ is the Fock state of the microwave field, we obtain:
\begin{equation}
H_0=\left(
\begin{array}{cc}
\omega_{10}(t)/2+n\omega & \Omega_R/2 \\
\Omega_R/2 &-\omega_{10}(t)/2+(n+1)\omega \\
\end{array}
\right) , \label{Hamiltonian0}
\end{equation}
where $\Omega _{R}=g\sqrt{n+1}$ is  the Rabi frequency.
$%
H_{0}$ can be transformed to:
\begin{equation}
H_0=\left(
\begin{array}{cccc}
0 & \Omega_R/2 \\
\Omega_R/2 & \delta (t) \\
\end{array}%
\right) , \label{Hamiltonian00}
\end{equation}%
where $\delta (t)=\omega -\omega _{10}(t)$ is the detuning. The coupling
between the two microwave dressed states $|0,n+1\rangle$ and $|1,n\rangle$ is analog to the
tunnel splitting between the $|$$\uparrow$$\rangle$ and $|$$\downarrow$$\rangle$ states of
a spin in the presence of a weak transverse magnetic field. The LZS interference occurs when the
triangle pulse sweeps back
and forth across this avoided crossing whose minimum gap is $\Omega _{R}$.
When the microwave power increases, $\Omega _{R}$ increases, which
subsequently affects the detailed structures but has negligible effect on the positions of
the fringes, as shown in Fig. 2(e). Here the one-dimensional data are
extracted from Fig. 2(a) ($\Omega _{Ra}/2\pi=19.6$ MHz, green
circles), 2(b) ($\Omega _{Rb}/2\pi=27.8$ MHz, red circles) and 2(c) ($\Omega _{Rc}/2\pi=41.7$ MHz, blue circles) at $\Phi _{LZS}=5$ m$\Phi _{0}$. Notice that all parameters used in the numerical simulations, such as Rabi frequencies and relaxation times of the qubit and TLSs, are obtained directly from experiments so that there is no fitting parameter. The
maximum height of the interference peaks was reached earlier with the
stronger microwave field, because a stronger microwave field leads to a larger splitting and thereby the phase difference accumulates at a faster rate.
Another interesting phenomenon in Fig. 2(e) is the reversal in the order of the peak height for
different microwave powers in the three periods of oscillations. They agreed with the calculated
results (color symbols in the inset of Fig. 2(e)) using the LZ transition theory in the $P_{T}-\eta$ plane, where $%
P_{T}(\eta)=4\exp (-\eta )(1-\exp (-\eta ))$, $\eta \equiv
2\pi(\Omega _{Ri}/2)^2/\nu$, $i=a, b, c$ and $\nu $ denotes the rate of the
changing energy level spacing of the noninteracting levels.

These results indicate that by adjusting the sweep rate of the triangle pulse,
the microwave power and frequency, one can control the qubit states coherently.

Though the LZS interferometry has been used mostly to characterize the
parameters defining the quantum systems and their interaction with the
environment, recent work suggests that it has great potential in the coherent manipulation of
quantum states, in particular multipartite quantum states \cite%
{J.R.Petta02052010,Nat.Commun.Sun}. In this context, the existence of avoided crossings
in the energy diagram of single qubit or coupled multiple qubits are crucial to producing the LZS interference. The
disadvantage of such intrinsic avoided crossings is that it is usually difficult to
control the location and the gap size \emph{in situ} once the qubits are fabricated.
A more fundamental problem is that for certain types of qubits
such as the SPQ, the computational basis states do not have intrinsic avoided crossings.
But in the LZS interferometry of the microwave dressed states as discussed above, one can create and/or adjust the
position and gap size of the avoided crossings as one desires.
This method is particularly advantageous in manipulating the states of multi-qubit systems as discussed below.

Note that in Fig. 2 when the tip of the triangle pulse reached the center of the qubit-TLS1
avoided crossing $\Phi _{TLS1}$, another group of interference fringes
emerged. These additional LZS interference fringes, which actually were similar to those observed in the previous work \cite{Nat.Commun.Sun}, were the results of the
coupling between the qubit and TLS1.
Taking into account the existence of TLS1, the Hamiltonian of the entire qubit-TLS-microwave filed system
becomes $H_{1}=H_{0}+H_{T_{1}}+H_{q-T_{1}}$. The Hamiltonian of TLS1 is $%
H_{T_{1}}=\frac{1}{2}\omega _{T_{1}}\sigma _{z}^{T_{1}}$, where $\omega
_{T_{1}}$ is the energy level spacing of TLS1 and $\sigma _{z}^{T_{1}}$ is Pauli Z operator on the TLS1.
The interaction Hamiltonian
is $H_{q-T_{1}}=g_{1}\sigma _{x}^{q}\otimes \sigma _{x}^{T_{1}}$, where $%
g_{1}$ is the coupling strength between the qubit and TLS1, $\sigma _{x}^{q}$ and $\sigma _{x}^{T_{1}}$
are Pauli X operators on the qubit and TLS1, respectively. In the subspace
spanned by $\{|1g,n\rangle ,|0g,n+1\rangle ,|0e,n\rangle \}$, $H_{1}$ can be simplified as:
\begin{equation}
H_1=\left(
\begin{array}{cccc}
0 & \Omega_R/2 & g_{1}\\
\Omega_R/2 & \delta(t) & 0\\
g_{1} &0 &\delta_{1}(t)\\
\end{array}%
\right) , \label{Hamiltonian1}
\end{equation}%
where $\delta _{1}(t)=\omega _{T1}-\omega _{10}(t)$. To facilitate quantitative comparisons between the theory and experiment, we averaged over different values of $n$ assuming
the microwave field is in a coherent state \cite{PhysRevLett.98.257003} characterized by $\langle
n\rangle $ and solved the corresponding Bloch
equation numerically. The results agree quantitatively with the experimental
data, as shown in Fig. 2.

Furthermore, the initial state of the qubit could dramatically affect the LZS interference of
the microwave dressed states of the SPQ.
Of special interest is when the qubit is initially biased at the point where the microwave is resonant with $%
\omega _{10}$, i.e., $\Phi _{i}$ = $\Phi _{r}$. In this case, we set $%
t_{mw}\geq t_{\Lambda}$ as shown with dotted line in Fig. 1(c) and found that the difference between
them, $t_{i}=t_{mw}-t_{\Lambda}$, affected the interference fringes significantly. We measured the LZS
interference for $t_{i}$$\ =\ $0, 7 ns, 13 ns, 19 ns and 750 ns,
corresponding to the 0, $\pi $/2 pulse, $\pi $ pulse, 3$\pi $/2 pulse and
mixed states in Rabi oscillation with $\omega /2\pi =16.345$ GHz and the
nominal power 13 dBm, respectively. As shown in Fig. 3(a)-3(e), the interference
fringes are very sensitive to $t_{i}$, because different $t_{i}$ results
in different initial states of the qubit at the beginning of the triangle pulse, i.e., different probability
amplitudes of $|0\rangle $ and $|1\rangle $. The corresponding numerical results are shown in the insets.
It is noticed that in Fig. 3(a), both $\Phi_{LZS}\geq\Phi_i$ and $\Phi_{LZS}<\Phi_i$ are included.
Thus three microscopic TLSs (TLS1, TLS2, and TLS3) were involved into the evolution as shown
in Fig. 1(a) and Fig. 1(c). The qubit and the three TLSs now form a tetrapartite
quantum system, the Hamiltonian of which is $H_{2}=H_{0}+\sum_{i=1}^{3}%
\frac{1}{2}\omega _{T_{i}}\sigma _{z}^{T_{i}}+\sum_{i=1}^{3}g_{i}\sigma
_{x}^{q}\otimes \sigma _{x}^{T_{i}}$. In the subspace spanned by $%
\{|1g_{1}g_{2}g_{3},n\rangle ,\ |0g_{1}g_{2}g_{3},n+1\rangle ,\
|0e_{1}g_{2}g_{3},n\rangle ,\ |0g_{1}e_{2}g_{3},n\rangle ,\
|0g_{1}g_{2}e_{3},n\rangle \}$, $H_{2}$ can be written as:
\begin{equation}
H_2=\left(
\begin{array}{ccccc}
0 & \Omega_R/2  & g_{1} & g_{2} & g_{3}\\
\Omega_R/2  & \delta(t) & 0 & 0 & 0\\
g_{1} &0 &\delta_{1}(t) & 0 & 0\\
g_{2} &0 & 0 &\delta_{2}(t) & 0 \\
g_{3} &0 & 0 & 0 &\delta_{3}(t) \\
\end{array}%
\right), \label{Hamiltonian2}
\end{equation}%
where $\delta _{i}(t)=\omega _{TLSi}-\omega _{10}(t)$, and $g_{i}$ is the
coupling strength between the qubit and the $i$th TLS ($i$=1, 2, 3). Considering that the simulation is done with no free parameters, the agreement between the numerical simulation and
experimental results presented in Fig. 3 is quite remarkable. It should be pointed
out that the width of the triangle pulse is much shorter than the decoherence time.
Therefore, the system's evolution remains coherent. To the best of our knowledge,
this is the first demonstration of controlling tetrapartite coherent evolution in a solid-state quantum system.

The overall good agreement not only confirms the validity of our understanding and treatment of
the multipartite system interacting with the microwave field, but further proves that the LZS interferometry
of the microwave dressed states can be a powerful tool for controlling multi-partite quantum systems and
enhancing the efficiency and flexibility of the novel approach based on LZS
interferometry \cite{Nat.Commun.Sun} to multi-qubit quantum gates.
Compared to the Stark-chirped rapid adiabatic passage method \cite%
{PhysRevLett.32.814,PhysRevLett.100.113601}, the quantum gates based on the LZS interferometry of the microwave dressed states achieve
two significant improvements: (i) The qubit flux bias remains unchanged after the gate
operations. This will significantly simplify the subsequent operations and increase the gate fidelity;
(ii) The location and intensity of
the LZS interference can be controlled \emph{in situ} by adjusting the
external parameters. It is thus very promising for quantum information applications such as the implementation of much faster
multi-qubit quantum gates.

Our experiment has verified that the macroscopic artificial atom, SPQ, interacts with the microwave
field in the same way as the atoms interact with the light. The concept of dressed states not only provides an
excellent intuitive picture to understand qualitatively the behavior of such complicated system but also
a theoretical foundation for quantitative simulation and prediction of the system's dynamics. It should be emphasized that this method of coherent state control of multipartite systems involving the dressed states is not limited to the system studied here which includes an artificial atom (the phase qubit) and several unintended microscopic TLSs. In fact, the method is applicable to ANY quantum systems which have avoided level crossings resulting from interaction between the individual constituents and photons. For instance, it can be applied to the coupled quantum dots, resonator coupled phase qubits, inductively coupled flux qubits, and nuclear and electron spins interacting with electromagnetic waves (e.g., lights, microwaves). Thus, the approach presented in this work can be generalized readily to other systems interacting with electromagnetic fields,
opening a new route toward the realization of large scale quantum information processing.

\begin{acknowledgments}
This work was supported in part by MOST (2011CB922104, 2011CBA00200), NCET, NSFC (11074114, 91021003), NSF (DMR-0325551) and DMEA (H94003-04-D-0004-0149).
We gratefully acknowledge Northrop Grumman ES in Baltimore MD for foundry support and
thank R. Lewis, A. Pesetski, E. Folk, and J. Talvacchio for technical assistance.
\end{acknowledgments}

\clearpage
\begin{figure}
\includegraphics[width=0.5\textwidth]{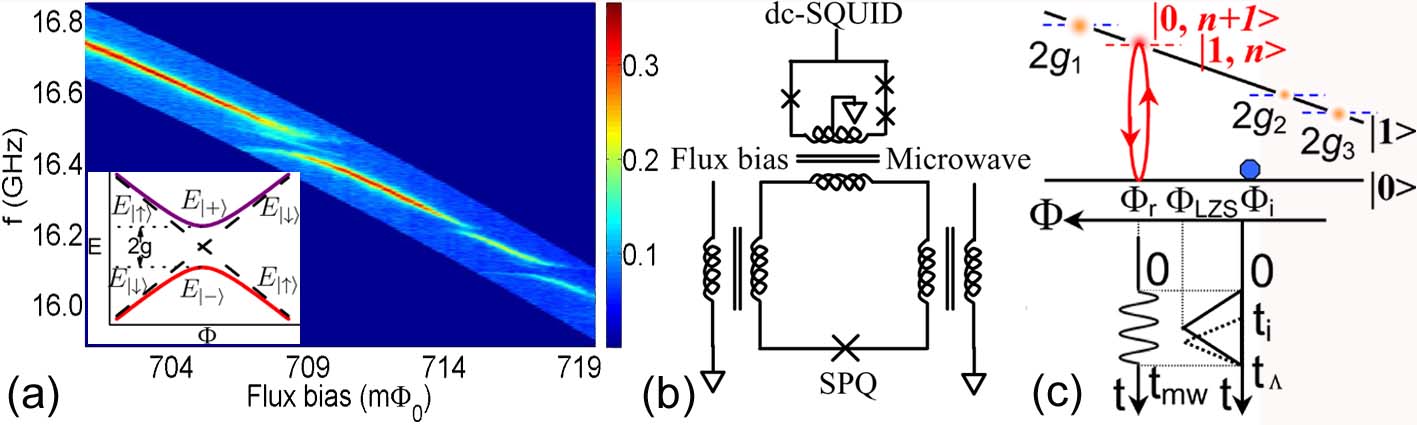}
\caption{(a)  Measured spectroscopy of the SPQ. The inset shows a general avoided level crossing. The dashed and solid lines represent the energies of the uncoupled and coupled states, respectively. (b) Schematic of the qubit circuitry. Detailed parameters are described in the previous work \cite{Nat.Commun.Sun}. (c) Schematic of measuring the LZS interference. The avoided crossings caused by the qubit-TLS interaction are also shown.}
\end{figure}

\begin{figure}
\includegraphics[width=0.5\textwidth]{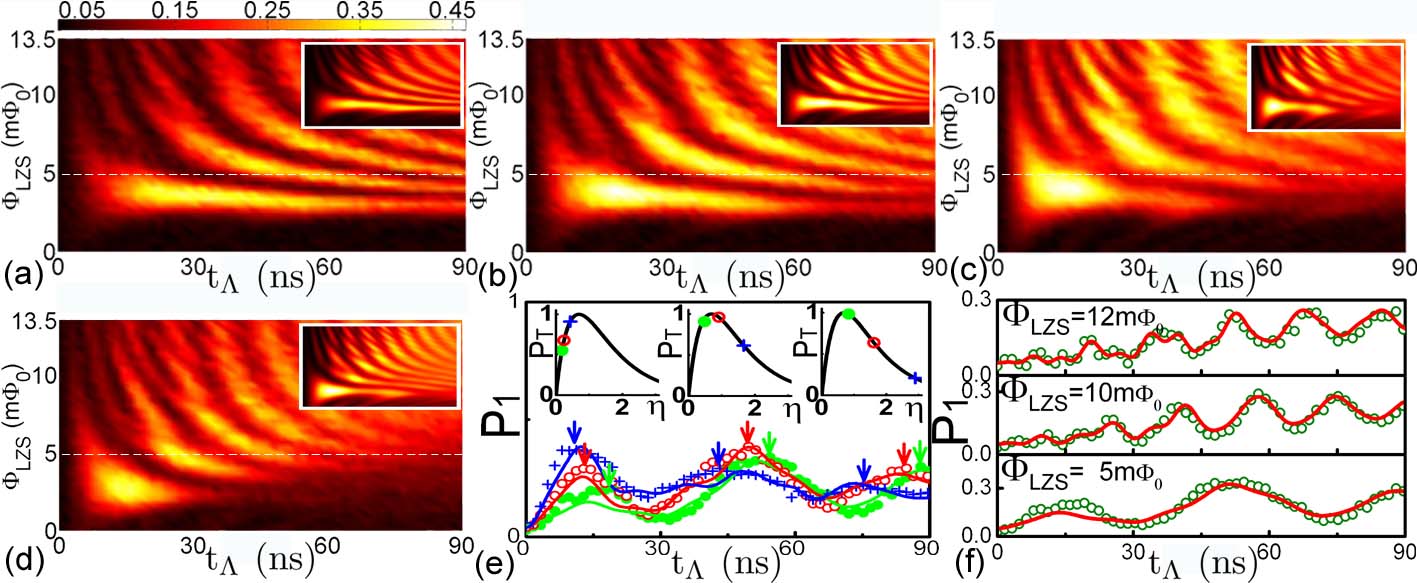}
\caption{(a, b, c) The LZS interference with
 $\omega/2\pi = 16.345$ GHz. The Rabi frequencies
 are $\Omega_{Ra}$, $\Omega_{Rb}$ and $\Omega_{Rc}$,
 respectively. The inset is the numerical result using $H_1$. (d) The LZS interference with $\omega/2\pi = 16.315$ GHz and Rabi
 frequency 30.9 MHz. (e) $P_1(t_{\Lambda})$ at $\Phi_{LZS}= 5 $ m$\Phi _{0}$
with $\Omega_{Ra}$ (green dots), $\Omega_{Rb}$ (red circles)
and $\Omega_{Rc}$ (blue crosses), respectively. Maxima are marked with color arrows.
The insets show the positions of each maximum with color symbols in $P_T(\eta)$. (f) One-dimensional experimental data (circles) and numerical ones (lines) extracted from Fig. 2(a). The positions of TLS are marked with the dashed lines.}
\end{figure}

\begin{figure}
\includegraphics[width=0.5\textwidth]{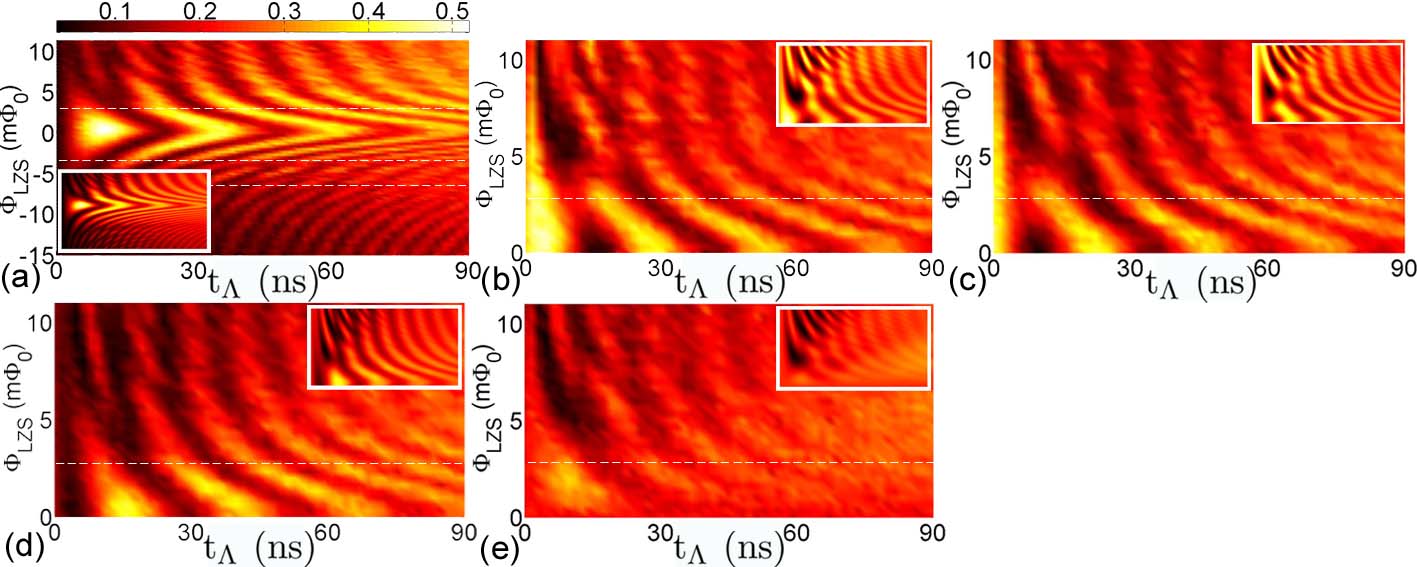}
\caption{Dependence of the LZS interference beween the microwave dressed states on the initial state of
the qubit as characterized by $t_{i}$  (c.f. Fig. 1(c)). (a) $t_{i}$=0 corresponds to the initial
state $\Psi_i=|0\rangle $. (b) $t_{i}$=7 ns, for $\Psi_i=(|0\rangle-i|1\rangle)/\sqrt{2}$.
(c) $t_{i}$=13 ns, for $\Psi_i=|1\rangle $. (d) $t_{i}$=19 ns,
for $\Psi_i=(|0\rangle+i|1\rangle)/\sqrt{2}$. (e) $t_{i}$=750 ns, for the mixed
state $\frac{1}{2}$($|0\rangle$$\langle 0|$+$|1\rangle $$\langle 1|$).
The insets are the numerically simulated results, in which the qubit's relaxation and dephasing time is 70 ns and 80 ns, respectively, and the TLS's decoherece time is 150 ns. The positions of TLS are marked with the dashed lines.}
\end{figure}


\begin{thebibliography}{1}

\bibitem{Shevchenko2010} S. Shevchenko \emph{et al}., Physics Reports \textbf{492}, 1 (2010).
\bibitem{PhysRevLett.69.1919} S. Yoakum \emph{et al}., Phys. Rev. Lett. \textbf{69}, 1919 (1992).
\bibitem{AnnalsofPhysics} T. D. Clark \emph{et al}., Annals of Physics \textbf{268}, 1 (1998).
\bibitem{WilliamD.Oliver12092005} W. D. Oliver \emph{et al}., Science \textbf{310}, 1653 (2005).
\bibitem{PhysRevLett.96.187002} M. Sillanp\"{a}\"{a} \emph{et al}., Phys. Rev. Lett. \textbf{96}, 187002 (2006).
\bibitem{PhysRevLett.98.257003} C. M. Wilson \emph{et al}., Phys. Rev. Lett. \textbf{98}, 257003 (2007).
\bibitem{PhysRevLett.101.017003} A. Izmalkov \emph{et al}., Phys. Rev. Lett. \textbf{101}, 017003 (2008).
\bibitem{PhysRevLett.101.190502} M. S. Rudner \emph{et al}., Phys. Rev. Lett. \textbf{101}, 190502 (2008).
\bibitem{sun:102502} G. Sun \emph{et al}., Appl. Phys. Lett. \textbf{94},102502 (2009).
\bibitem{Nature.459.960} M. D. LaHaye \emph{et al}., Nature \textbf{459}, 960 (2009).
\bibitem{Nat.Commun.Sun} G. Sun \emph{et al}., Nature Communications \textbf{1}, 51 (2010).
\bibitem{J.R.Petta02052010} J. R. Petta \emph{et al}., Science \textbf{327}, 669 (2010).
\bibitem{Garg} A. Vijayaraghavan and A. Garg, Phys. Rev. B \textbf{79}, 104423 (2009).
\bibitem{API} C. Cohen-Tannoudji \emph{et al}., Atom-Photon Interactions: basic processes and applica-
tions (Wiley, New York, 1992).
\bibitem{RevModPhys.73.357} Y. Makhlin \emph{et al}., Rev. Mod. Phys. \textbf{73}, 357 (2001).
\bibitem{PhysicsToday.58.11} J. You and F. Nori, Physics Today \textbf{58}, 42 (2005).
\bibitem{Nature.453.1031} J. Clarke and F. K. Wilhelm, Nature \textbf{453}, 1031 (2008).
\bibitem{PhysRevA.67.042311} C.-P. Yang \emph{et al}., Phys. Rev. A \textbf{67},
042311 (2003).
\bibitem{PhysRevB.68.064509} J. Q. You and F. Nori, Phys. Rev. B \textbf{68}, 064509 (2003).
\bibitem{Nature451.664} R. J. Schoelkopf and S. M. Girvin, Nature \textbf{451}, 664
(2008).
\bibitem{PhysRevB.81.024520} C. M. Wilson \emph{et al}., Phys. Rev. B \textbf{81}, 024520
(2010).
\bibitem{QuantumInfProcess.8.81} J. M. Martinis, Quantum Inf. Process. \textbf{8}, 81 (2009).
\bibitem{QuantumInfProcess.8.117} R. W. Simmonds \emph{et al}., Quantum Inf. Process. \textbf{8}, 117
(2009).
\bibitem{NaturePhysics.4.523} M. Neeley \emph{et al}., Nature Physics \textbf{4}, 523 (2009).
\bibitem{PhysRevLett.97.077001} A. M. Zagoskin \emph{et al}., Phys. Rev. Lett. \textbf{97}, 077001 (2006).
\bibitem{PhysRevLett.32.814} M. M. T. Loy, Phys. Rev. Lett. \textbf{32}, 814 (1974).
\bibitem{PhysRevLett.100.113601} L. F. Wei \emph{et al}., Phys. Rev. Lett. \textbf{100}, 113601 (2008).

\end{thebibliography}
\end{document}